\documentclass[prl,twocolumn,showpacs,floatfix]{revtex4}
\usepackage{graphicx}
\usepackage{amsmath}

\def\RR{{\rm\kern.24em \vrule width.04em height1.46ex depth-.07ex
\kern-.30em R}}

\newcommand{\ket}[1]{\left \vert #1 \right \rangle}
\newcommand{\bra}[1]{\left \langle #1 \right \vert}

\begin{document}

\title{Dressed Qubits}
\author{L.-A. Wu and D. A. Lidar}
\affiliation{Chemical Physics Theory Group, University of Toronto, 80 St. George St.,
Toronto, Ontario M5S 3H6, Canada}

\begin{abstract}
Inherent gate errors can arise in quantum computation when the actual system
Hamiltonian or Hilbert space deviates from the desired one. Two important
examples we address are spin-coupled quantum dots in the presence of
spin-orbit perturbations to the Heisenberg exchange interaction, and
off-resonant transitions of a qubit embedded in a multilevel Hilbert space.
We propose a ``dressed qubit'' transformation for dealing with such inherent
errors. Unlike quantum error correction, the dressed qubits method does not
require additional operations or encoding redundancy, is insenstitive to
error magnitude, and imposes no new experimental constraints.
\end{abstract}

\maketitle

In theoretical models of physical systems implementing quantum computers it
is common to make simplifying assumptions about the form of the underlying
system Hamiltonian, or the system's interaction with external controls. Such
assumptions lead to an idealized system description that is convenient for
the purpose of proving desirable properties, such as the ability to perform
universal quantum computation (QC) \cite{Nielsen:book}. Important examples
of simplifying assumptions include the neglect of certain interactions
(e.g., of spin-orbit coupling in models of quantum-dot quantum computers 
\cite{Loss:98,Kavokin:0102}), and reduction of a multi-level Hilbert space to
a two-dimensional one (thus neglecting off-resonant effects, e.g., in
superconducting qubits \cite{Mooij:99,Tian:00,Yu:02Blais:02}). Approaches for
dealing with the actual system, $S_{A}$, as opposed to the idealized system, 
$S_{I}$, typically (though not always \cite{WuLidar:02}) treat the
difference $\Delta =S_{A}-S_{I}$ between the two as a \emph{problem} that
needs to be overcome:\ $\Delta $ is considered an \emph{inherent error}. For
example, shaped pulses have been proposed to correct for the inevitable
appearance of spin-orbit corrections in quantum dots \cite{Bonesteel:01},
quantum error correcting codes have been shown to correct certain systematic
qubit-qubit interaction errors \cite{Gea:00}, a problem also approached
using NMR-inspired composite pulse sequences \cite{Cummins:02}, and
off-resonant transitions have been shown to be cancellable via a sequence of
resonant pulses \cite{Tian:00} or optimal control fields \cite{Palao:02}.
The motivation for removing $\Delta $ typically comes from the fact that
there already exists a theory enabling convenient universal QC using $S_{I}$, but the same theory does not apply using $S_{A}$. For example, in
interacting spin systems, such as quantum dots, it is known how to perform
universal QC by manipulating only isotropic Heisenberg exchange interactions 
\cite{Bacon:99aKempe:00,DiVincenzo:00a}, an attractive prospect that
eliminates the need for performing difficult single-qubit operations. In
this case $S_{I}$ is the purely isotropic Heisenberg Hamiltonian and $\Delta 
$ is the spin-orbit correction. The presence of $\Delta $ spoils the just
mentioned universality result, and hence methods to cancel $\Delta $ have
been proposed \cite{Bonesteel:01,Burkard:01}. Is it always necessary to
cancel $\Delta $? Here we introduce a new method, termed ``dressed
qubits'', that enables universal QC using $S_{A}$, provided
it is known how to perform universal QC using $S_{I}$. This includes
preparation and measurement of quantum information, in the dressed qubit basis. Thus, the ability to,
e.g., perform universal QC using only Heisenberg interactions, translates
directly into the ability to perform universal QC under Heisenberg
interactions including spin-orbit coupling $\Delta $, \emph{without any
extra overhead, and irrespective of the magnitude of }$\Delta $. These are
general features of the dressed qubits method, that distinguish it from all
schemes trying to correct for $\Delta $, as opposed to working directly with 
$S_{A}$. The underlying idea is to find a unitary \textquotedblleft
dressing'' transformation between the ``ideal qubit basis''
(the one for which universality results can be proven relatively easily) and
the ``dressed qubit basis'' (corresponding to computation using $S_{A}$).
We first give a general description of the dressed qubit method. We then
illustrate the general results with examples of relevance to promising QC
proposals.

\textit{Generalities}.--- Suppose that a
system $S_{A}$ of $N$ (physical or encoded) qubits possesses a set of 
experimentally controllable Hamiltonians (or corresponding
evolution operators) $\mathbf{H}=\{H_{\alpha }\}$ (or $\mathbf{U}(\mathbf{
\theta })=\{U_{\alpha }=e^{-i\theta _{\alpha }H_{\alpha }}\}$), which may be
accompanied by inherent errors $\Delta $. Correspondingly,
there is an idealized set of Hamiltonians (or corresponding
evolution operators) $\mathbf{H}^{\rm id}=\{H_{\alpha }^{\rm id}\}$ (or $\mathbf{U}
^{\rm id}(\mathbf{\theta }^{\rm id})=\{U^{\rm id}=e^{-i\theta _{\alpha }^{\rm id}H_{\alpha
}^{\rm id}}\}$), which is universal: the set $\mathbf{H}^{\rm id}$ can be used to
generate a transformation between an arbitrary $N$-qubit state $\ket{\Psi ^{\rm id}}$ and any other such state. Assume that there exists
a \emph{fixed} unitary transformation $V$ such that $H_{\alpha
}=V^{\dagger }H_{\alpha }^{\rm id}V$ $\forall \alpha$. Then we define $V$ as the ``dressing
transformation'' between $\mathbf{H}^{\rm id}$ and $\mathbf{H}$,
and the states $\ket{\Psi } =V^{\dagger }\ket{\Psi
^{\rm id}} $ (where $\ket{\Psi ^{\rm id}} $ is an arbitrary 
$N$-qubit state of the idealized, or ``bare'' system $S_{I}$) and $\ket{\Psi
} $ are the ``dressed states''.
It follows that for all $|\Phi^{({\rm id})}\rangle,|\Psi^{({\rm id})}\rangle$
\begin{equation}
\bra{\Phi} H_{\alpha }\ket{\Psi }
=\bra{\Phi ^{\rm id}} H_{\alpha }^{\rm id}\ket{\Psi
^{\rm id}} ,  \label{eq:matelems}
\end{equation}
i.e., matrix elements in the dressed basis are identical to those in the
idealized basis. Hence QC in the idealized and dressed (actual) system is
equivalent. \emph{The dressing transformation $V$ need not be
implementable experimentally}. We do, however, require that states can be
prepared and measured in the dressed basis, so that this basis can be used
for input and output. $V$ may be separable: $V=\bigotimes_{j=1}^{N}V_{j}$.
In this case, one can specifically define a dressed qubit represented by
states $\ket{0} =V^{\dagger }\ket{0^{\rm id}} 
$ and $\ket{1} =V^{\dagger }\ket{1^{\rm id}} $
, where $\ket{0^{\rm id}} $ and $\ket{1^{\rm id}} $
are the ``bare'', or idealized
computational basis states. Such a separable dressing transformation retains
essentially the features of $\mathbf{H}$ when it is transferred into $
\mathbf{H}^{\rm id}$, meaning that a one (two)-qubit operation in $\mathbf{H}$ is
transferred into the corresponding one (two)-qubit operation in $\mathbf{H}
^{\rm id}$. Below we discuss both separable and non-separable dressing
transformations.

While the notion of dressed qubits is simple, it is usually not
straightforward to find a valid $V$ for a particular physical system,
since one has to consider both single- \emph{and} two-qubit operations in
order for the general relation $H_{\alpha
}=V^{\dagger }H_{\alpha }^{\rm id}V$ to hold. Surprisingly, we
report here that some promising QC proposals with inherent errors are
dressable.

Below we make repeated use of the following identity, valid for any set of
operators $\{J_{x},J_{y},J_{z}\}$ satisfying the two su$(2)$ [or
so$(3)$] commutation relations $[J_{z},J_{x}]=iJ_{y}$,
$[J_{y},J_{z}]=iJ_{x}$ (the third relation $[J_{x},J_{y}]=iJ_{z}$ is
not required):
\begin{equation}
\sqrt{1+\delta ^{2}} e^{-i\varphi J_{z}}J_{x} e^{i\varphi J_{z}}=J_{x}+\delta J_{y},\,\,\, \delta =\tan \varphi .  \label{eq:key}
\end{equation}

\textit{Eliminating off-resonant effects.---} In many QC proposals a
two-dimensional qubit subspace is embedded in a larger $N$-level Hilbert
space. In such cases quantum logic operations typically mix the qubit
subspace with the other states. This is known as \textquotedblleft
leakage'', and is the result of unwanted off-resonant
transitions \cite{Tian:00,Palao:02}. Since it follows from
time-independent perturbation
theory that such transitions are stronger for levels closer to those
supporting the qubit, we consider for simplicity first a three-level model
with states $\{\ket{k^{\rm id}} \}_{k=0}^{2}$. The first two
states represent the qubit. This example is highly relevant to
superconducting QC proposals (e.g., the current-biased Josephson
junction \cite{Yu:02Blais:02}, and the persistent-current qubit \cite{Mooij:99}),
where the qubit levels couple to a third level supported by the potential.
Ideally $\{H_{1}^{\rm id}=f\sqrt{1+\delta ^{2}}
(c_{0}^{\dagger }c_{1}+c_{1}^{\dagger }c_{0}),H_{2}^{\rm id}=\epsilon n_{1}\equiv
\epsilon c_{1}^{\dagger }c_{1}\}$, where $c_{k}^{\dagger }$ is a fermionic
or bosonic creation operator for level $k$, or, in the case of a
single-particle Fock space, a projection operator
such that $c_{k}^{\dagger }c_{l} = \ket{k}\bra{l}$. The representations of $H_{1}^{\rm id}$ and $H_{2}^{\rm id}$ in
the two-dimensional qubit subspace are the Pauli matrices $\sigma _{x}$ and $
\sigma _{z}$, respectively, and generate an SU$(2)$ group for all
single-qubit operations. Experimentally, instead one typically obtains the
actual Hamiltonian $H_{1}=f[(c_{0}^{\dagger }c_{1}+c_{1}^{\dagger
}c_{0})+\delta (c_{1}^{\dagger }c_{2}+c_{2}^{\dagger }c_{1})]$, where the
last term is the undesirable off-resonant transition. Additionally, we
now have $H_{2}= \sum_{i=1}^2 \epsilon_i n_{i}$, where we assume that
$\epsilon_1 \neq \epsilon_2$ are tunable, or else rational multiples of each other. Effective, but
costly schemes have been proposed to eliminate the systematic error due to $
\delta $ \cite{Tian:00,Palao:02}.
As an alternative to
eliminating $\delta $, using a dressed qubit instead of $\ket{0^{\rm id}} $ and $\ket{1^{\rm id}} $ solves the problem
at no extra cost: First, note that the set $\{X\equiv c_{0}^{\dagger
}c_{1}+c_{1}^{\dagger }c_{0},Y\equiv c_{1}^{\dagger }c_{2}+c_{2}^{\dagger
}c_{1},Z\equiv i(c_{2}^{\dagger }c_{0}-c_{0}^{\dagger }c_{2})\}$
satisfies su$(2)$ or so$(3)$ commutation relations. Hence a
possible dressing transformation for the $k$th qubit is $V_{k}=\exp (\varphi
_{k}(c_{2}^{\dagger }(k)c_{0}(k)-c_{0}^{\dagger }(k)c_{2}(k))$, where $
\varphi _{k}=\tan ^{-1}\delta _{k}$. Using Eq.~(\ref{eq:key}), it follows
that for a dressed qubit $\ket{\Phi } _{k}=V_{k}\ket{\Phi ^{\rm id}} _{k}$ (a superposition of the states $\ket{0^{\rm id}}_k $ and $\ket{2^{\rm id}}_k $), $
H_{1}=f(X+\delta Y)$ acts as $\sigma _{x}$: 
\begin{eqnarray*}
\bra{\Psi} H_1 \ket{\Phi } = 
\bra{\Psi ^{\rm id}}f\sqrt{1+\delta
  ^{2}}(c_{0}^{\dagger}c_{1}+c_{1}^{\dagger }c_{0})\,\ket{\Phi ^{\rm
    id}} .
\end{eqnarray*}
$H_2$ no longer acts as $\sigma _{z}$ in the presence of $n_2$, since
$[H_2(k),V_k] \neq 0$. However, as long as $\epsilon_1 \neq \epsilon_2$ are
rationally related or tunable, by letting the system evolve under $H_2$ for an
appropriate duration, it is always possible to effectively cancel
$n_2$ by evolving it for time $2\pi/\epsilon_2$,
while letting $n_1$ at the same time generate one of the discrete set of single-qubit
quantum gates that are known to be universal together with $\sigma
_{x}$ and an entangling two-qubit gate \cite{Nielsen:book}.
The dressing transformation is compatible,
e.g., with the two-qubit Ising-like interaction $n_{1}(i)n_{1}(j)$, since
then $n_{1}(k)n_{1}(l)=V_{k}^{\dagger }V_{l}^{\dagger
}n_{1}(k)n_{1}(l)V_{l}V_{k}$. In general, if the single qubit operations undergo the transformation $V_{k}$, the two-qubit interaction undergoes the transformation $V_{k}V_{l}$. This
transformation must ensure that the actual two-qubit interaction
becomes the idealized two-qubit interaction. In our discussion of
exchange interactions below we give a non-trivial such example.

The construction above is easily generalized to systems with $N>3$ levels
with an interaction of the form $H_{1}^{(N)}=f[c_{0}^{\dagger
}c_{1}+\sum_{j=2}^{N-1}\delta _{j}c_{1}^{\dagger }c_{j}+h.c.]$, describing
leakage from state $\ket{1^{\rm id}} $ to all states $
\{\ket{j^{\rm id}} \}_{j=2}^{N-1}$. Let $|\kappa
|^{2}=\sum_{j=2}^{N-1}|\delta _{j}|^{2}$; the triple $\{X\equiv
c_{0}^{\dagger }c_{1}+c_{1}^{\dagger }c_{0},Y\equiv \frac{1}{|\kappa |}
\sum_{j=2}^{N-1}(\delta _{j}c_{1}^{\dagger }c_{j}+h.c.),Z\equiv \frac{i}{
|\kappa |}\sum_{j=2}^{N-1}(\delta _{j}c_{j}^{\dagger }c_{0}-h.c.)\}$
is an su$(2)$ or so$(3)$ algebra. Therefore $H_{1}^{(N)}=f(X+|\kappa | Y)=f\sqrt{1+|\kappa |^{2}}
e^{-i\varphi Z}Xe^{i\varphi Z}$, where $\varphi =\tan ^{-1}|\kappa |$. The
general-$N$ dressing transformation is thus $V=\exp [\frac{\varphi }{
|\kappa |}\sum_{j=2}^{N-1}(\delta _{j}c_{j}^{\dagger }c_{0}-h.c.)]$, which
creates a dressed qubit that is a superposition of states $\ket{
0^{\rm id}} $ and $\{\ket{j^{\rm id}} \}_{j=2}^{N-1}$,
and is again compatible with the Ising interaction.

Now note that $\ket{1} =\ket{1^{\rm id}} $. Therefore \emph{preparation}
amounts to initializing all qubits in the state $\ket{
1^{\rm id}} $, and \emph{measurement in the dressed basis} amounts to
observing just the $\ket{1^{\rm id}} $ state. This can be done
similarly to the technique of cycling transitions in trapped ions, by
coupling the $\ket{1^{\rm id}} $ state to an auxiliary level
and observing fluorescence \cite{Steane:00}.

The dressed qubit is the natural computational basis given the actual
``leaky'' interaction $H_{1}$, and there is
no need to eliminate the ``leakage'' term
contained in $H_{1}$: this term represents leakage only with respect to the
``unnatural'' computational basis $\ket{0^{\rm id}} ,\ket{
1^{\rm id}} $. The dressed qubit is ``natural'' in the sense that there is no need to physically
implement the dressing transformation: it is inherent in the actual
Hamiltonian.

\textit{Encoded QC using Heisenberg interaction with anisotropy.---} The
Heisenberg exchange interaction $J\mathbf{S}_{k}\cdot \mathbf{S}_{l}$
between spins $\mathbf{S}_{k}$ and $\mathbf{S}_{l}$ is central to a number
of the most promising solid-state QC proposals, including electrons in
quantum dots \cite{Loss:98} and donor atoms in Si arrays \cite{Kane:00}. It
has been shown to be universal for QC, without (more difficult to implement)
single-qubit gates, provided one encodes a logical qubit into the state of
several spins \cite{Bacon:99aKempe:00,DiVincenzo:00a}. In reality, the
idealized Heisenberg Hamiltonian is, however, perturbed by an 
anisotropic term arising due to spin-orbit interactions: the actual
Hamiltonian is 
\begin{equation}
H_{kl}=J\{\mathbf{S}_{k}\cdot \mathbf{S}_{l}+\mathbf{D}\,\cdot \mathbf{S}
_{k}\times \mathbf{S}_{l}+\gamma (\mathbf{S}_{k}\cdot
\mathbf{D)(S}_{l}\cdot \mathbf{D)\},}
\label{eq:Hkl}
\end{equation}
where $\mathbf{D}\in {\RR}^{3}$ is known as the Dzyaloshinski-Moriya vector in
solid-state physics, and $\gamma =
\frac{\sqrt{1+|\mathbf{D}|^{2}}-1}{|\mathbf{D}|^{2}}$. Kavokin has
estimated that $|\mathbf{D}|$ is in the range $0.01-0.8$ in coupled quantum
dots in GaAs \cite{Kavokin:0102}. This is at least two orders of
magnitude beyond the current fault-tolerance threshold estimates of quantum
error correction theory \cite{Steane:02}. For this reason the anisotropic
perturbation has been considered a problem and strategies have been designed
to cancel it. E.g., it can be removed to first order by shaped pulses \cite
{Bonesteel:01}, cancelled in the absence of an external magnetic field and
in the presence of single-qubit operations \cite{Burkard:01}, or used in
order to generate a universal gate set that, however, incurs some timing
overhead \cite{WuLidar:02}. These approaches to dealing with the spin-orbit
term are motivated by universal QC with either the usual (bare) choice
of $S_{z}$ eigenstates as qubits \cite{Burkard:01,WuLidar:02}, or with
encoded qubits \cite{Bonesteel:01}. Here we show that \emph{dressed qubits,
defined with respect to the actual Hamiltonian} $H_{kl}$, \emph{offer a
solution that is fully compatible with the encoded qubits approach, at no
extra overhead and without any approximations, other than the assumption that} $\mathbf{D}$ \emph{is time-independent}. The residual time-dependence of $\mathbf{D}
$ (that arises via the spin-orbit constant from switching of $J$ during the
execution of quantum gates \cite{Kavokin:0102}) is small enough that it can be
corrected using QECC \cite{Burkard:01}.

We derive a dressing transformation by constructing a set of su$(2)$
operators for $H_{kl}$ of Eq.~(\ref{eq:Hkl}). The operators $\{X_{kl}\equiv \mathbf{S}_{k}\cdot 
\mathbf{S}_{l}-(\mathbf{S}_{k}\cdot \mathbf{n)(S}_{l}\cdot \mathbf{n),}
Y_{kl}\equiv \mathbf{n}\cdot (\mathbf{S}_{k}\times \mathbf{S}
_{l}),Z_{kl}\equiv \frac{1}{2}\mathbf{n}\cdot (\mathbf{S}_{l}-\mathbf{S}
_{k})\}$, where $\mathbf{n}$ is a unit vector, form such a set. Further note
that $[Z_{kl},(\mathbf{S}_{k}\cdot \mathbf{n})(\mathbf{S}_{l}\cdot \mathbf{
n})]=0$. It therefore follows by direct substitution from Eq.~(\ref{eq:key})
that 
\begin{equation}
W_{kl}=e^{-i\frac{1}{2}\epsilon \mathbf{n}\cdot (\mathbf{S}_{k}-\mathbf{S}
_{l})}  \label{eq:Vij}
\end{equation}
is a transformation such that $H_{kl}=W_{kl}^{\dagger
}H_{kl}^{\rm id}W_{kl}$, 
where $H_{kl}^{\rm id}=\sqrt{1+|\mathbf{D}|^{2}}J\mathbf{S}_{k}\cdot \mathbf{S}
_{l}$ is the isotropic Heisenberg interaction, $\epsilon =\tan
^{-1}|\mathbf{D}|$, and $\mathbf{n=D
}/|\mathbf{D}|$. Alternatively, the set $\{X_{kl}
\mathbf{,}Y_{kl},Z_{l}\equiv \mathbf{n}\cdot \mathbf{S}_{l})\}$ satisfies
the pair of su$(2)$ commutation relations $[Z_{l},X_{kl}]=iY_{kl}$, $
[Y_{kl},Z_{l}]=iX_{kl}$ (but $[X_{kl},Y_{kl}]\neq iZ_{l}$). It again follows
from Eq.~(\ref{eq:key}) that $V_{l}=e^{i\epsilon
  \mathbf{n}\cdot \mathbf{S}_{l}}$ is a transformation such that 
\begin{equation}
H_{kl}=V_{l}^{\dagger }H_{kl}^{\rm id}V_{l}=V_{k}H_{kl}^{\rm id}V_{k}^{\dagger }.
\label{eq:VHV2}
\end{equation}

Let us now recall the encoding under which the Heisenberg interaction
becomes universal for QC. The most economical encoding uses the two total
spin $S=1/2$ representations of three spin-$1/2$ particles to encode a qubit 
\cite{Knill:99a}. A convenient choice of encoded qubit basis states are the
two states: $|0_{L}^{\rm id}\rangle _{z}=|s\rangle _{12}|\uparrow
\rangle _{3}$, and $|1_{L}^{\rm id}\rangle _{z}=\sqrt{2/3}|\uparrow \rangle _{1}|\uparrow
\rangle _{2}|\downarrow \rangle _{3}-\sqrt{1/3}|t\rangle _{12}|\uparrow
\rangle _{3}$, where $|s\rangle _{12}=(|\uparrow \rangle _{1}|\downarrow
\rangle _{2}-|\downarrow \rangle _{1}|\uparrow \rangle _{2})/\sqrt{2}$ and $
|t\rangle _{12}=(\ket{\uparrow}_{1} \ket{\downarrow}_{2} + \ket{\downarrow}_{1}\ket{\uparrow}_{2})/\sqrt{2}$ are the singlet and triplet
states of spins $1,2$, respectively. The $z$ subscript indicates that
these two states have total spin projection $S_{z}=+1/2$. Because $H_{kl}^{\rm id}
$ is a scalar of total spin, a qubit can also be represented by states with
quantization axis along an arbitrary direction $\mathbf{n}$; in this case we
use the (obvious)\ notation $\ket{0_{L}^{\rm id}} _{\mathbf{n}
},\ket{1_{L}^{\rm id}} _{\mathbf{n}}$, and write an arbitrary
encoded qubit state as $\ket{\Phi ^{\rm id}} _{l}=a\ket{
0_{L}^{\rm id}} _{\mathbf{n}l}+b\ket{0_{L}^{\rm id}} _{
\mathbf{n}l}$ ($|a|^{2}+|b|^{2}=1$). In \cite{DiVincenzo:00a} a convenient
set of universal gates was found for the $|0_{L}^{\rm id}\rangle
_{z},|1_{L}^{\rm id}\rangle _{z}$ encoding: sequences chosen from $\mathbf{U}
^{\rm id}=\{U_{12}^{\rm id}(\theta ),U_{23}^{\rm id}(\theta ),U_{45}^{\rm id}(\theta
),U_{56}^{\rm id}(\theta ),U_{34}^{\rm id}(\theta )\}$ are universal for two qubits
encoded into the states of spins $1-3$ and $4-6$, respectively, where $
U_{kl}^{\rm id}(\theta )=\exp (-i\theta \mathbf{S}_{k}\cdot \mathbf{S}_{l})$. The
first four gates serve as logical single-qubit operations for the two
encoded qubits; the last operation, $U_{34}^{\rm id}(\theta )$, serves to
entangle the two encoded qubits via a controlled-phase ($CZ$) gate \cite{Nielsen:book}. Let us now show how to construct logic gates directly in
terms of the actual interaction $H_{kl}$.

The $l$th logical qubit is encoded by physical qubits $3l-2,3l-1,3l$. We
define an arbitrary $l$th dressed qubit by 
\begin{equation}
\ket{\Phi } _{l}=V_{3l-2,3l}^{\dagger }\ket{\Phi
^{\rm id}} _{l},  \label{eq:Phi_j}
\end{equation}
where $V_{3l-2,3l}$ is the dressing transformation with $V_{kl}=(W_{kl})^2$ as given in Eq.~(\ref{eq:Vij}). Consider how \emph{single-qubit
operations} act on this dressed qubit. Let $U_{kl}(\theta )=\exp (-i\theta
H_{kl})$. It follows from Eq.~(\ref{eq:VHV2}) that $U_{12}(\theta
)\ket{\Phi } _{1}=[V_{1}U_{12}^{\rm id}(\theta )V_{1}^{\dagger
}][V_{13}^{\dagger }\ket{\Phi ^{\rm id}} _{1}]=V_{13}^{\dagger
}U_{12}^{\rm id}(\theta )\ket{\Phi ^{\rm id}} _{1}$, and similarly $
U_{23}(\theta )\ket{\Phi } _{1}=V_{13}^{\dagger
}U_{23}^{\rm id}(\theta )\ket{\Phi ^{\rm id}} _{1}$. Therefore $
_{1}\bra{\Psi}U_{12(23)}(\theta )\ket{\Phi }
_{1}= \,_{1}\bra{\Psi ^{\rm id}}V_{13}V_{13}^{\dagger }U_{12(23)}^{\rm id}(\theta
)\ket{\Phi ^{\rm id}} _{1}= \,_{1}\bra{\Psi
^{\rm id}}U_{12(23)}^{\rm id}(\theta )\ket{\Phi ^{\rm id}} _{1}$, meaning
that matrix elements of $U_{12}(\theta )$ and $U_{23}(\theta )$ in the
dressed basis are identical to those in the idealized basis. Thus all single
encoded-qubit operations can be performed using $H_{kl}$, provided the
dressed basis is used.

\begin{figure}[tbp]
\hspace{1.5cm} \includegraphics[height=10cm,angle=270]{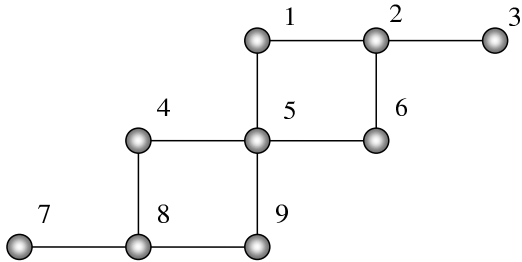} \vspace{-4cm
}
\caption{Geometry for 3-spin encoding. Each row represents a single
  encoded qubit.}
\label{fig1}
\end{figure}

Now consider \emph{two-qubit operations}. First, by using a sequence of
swaps, $U_{kl}^{{\rm id}\dagger }(\frac{\pi }{4}
)U_{lm}^{\rm id}(\theta )U_{kl}^{\rm id}(\frac{\pi }{4})=U_{km}^{\rm
  id}(\theta )$, we can replace the entangling gate $U_{34}^{\rm id}(\theta )$ by $
U_{15}^{\rm id}(\theta )$ or $U_{26}^{\rm id}(\theta )$. If we arrange the physical qubits as shown in Fig.~\ref
{fig1}, $U_{15}^{\rm id}(\theta )$ is a nearest neighbor interaction. Next, using
Eq.~(\ref{eq:Phi_j}), a two-encoded-qubit dressed state is $\ket{\Phi
} _{1}\ket{\Phi } _{2}=V_{13}^{\dagger
}V_{46}^{\dagger }\ket{\Phi ^{\rm id}} _{1}\ket{\Phi
^{\rm id}} _{2}$. We then have $U_{15}(\theta )\ket{\Phi
} _{1}\ket{\Phi } _{2}=[V_{1}U_{15}^{\rm id}(\theta
)V_{1}^{\dagger }][V_{13}^{\dagger }V_{46}^{\dagger }\ket{\Phi
^{\rm id}} _{1}\ket{\Phi ^{\rm id}}
_{2}]=V_{13}^{\dagger }V_{46}^{\dagger }U_{15}^{\rm id}(\theta )\ket{\Phi
^{\rm id}} _{1}\ket{\Phi ^{\rm id}} _{2}$, meaning that $
U_{15}(\theta )$ plays the same role in the dressed basis as does $
U_{15}^{\rm id}(\theta )$ in the idealized basis. Therefore the set $\mathbf{U}
=\{U_{12}(\theta ),U_{23}(\theta ),U_{45}(\theta ),U_{56}(\theta
),U_{15}(\theta )\}$ is universal for dressed qubits and has the same matrix
representations as in the idealized basis. With the arrangement shown in
Fig.~\ref{fig1}, spins $15,26,48,59,...$ are nearest neighbors, and $H_{kl}$
interactions between them can be used to generate a $CZ$ gate between any
pair of encoded qubits.

Next, we need to show that dressed qubits can be prepared and measured. Both
can be performed in a manner analogous to the procedure proposed in \cite{DiVincenzo:00a} for the idealized Heisenberg Hamiltonian. In that case the
computational basis state $\ket{0_{L}} _{\mathbf{n}
}=|s\rangle _{12}\ket{\uparrow } _{3\mathbf{n}}$ can be
prepared by turning on a strong exchange interaction between spins $1,2$,
and a moderately strong magnetic field $B\mathbf{n}$ (such that $k_{b}T\ll
g\mu _{B}B<J$): the system then relaxes to the ground state $|s\rangle _{12}
$ and spin $3$ is oriented along $\mathbf{n}$. The dressed state $\ket{
0_{L}} =V_{13}^{\dagger }|s\rangle _{12}\ket{\uparrow
} _{3\mathbf{n}}\propto e^{i\epsilon \mathbf{n\cdot S}
_{1}}|s\rangle _{12}\ket{\uparrow } _{3\mathbf{n}}$ can be
similarly prepared since it follows that $V_{13}^{\dagger }|s\rangle _{12}$
is the ground state of the actual Hamiltonian $H_{12}=V_{13}^{\dagger
}H_{12}^{\rm id}V_{13}$. Computation can then begin, with gates applied from the
set $\mathbf{U}$. The measurement scheme in \cite{DiVincenzo:00a} relies on
distinguishing a singlet $|s\rangle _{12}$ from a triplet $|t\rangle _{12}$
(e.g., using Kane's a.c. capacitance scheme \cite{Kane:00}), since this is a
measurement of whether the encoded qubit is in the state $|0_{L}\rangle
_{z}=|s\rangle _{12}|\uparrow \rangle _{3}$ or not (thus the state of spin $3
$ does not enter). In essence this is a measurement of the idealized
observable $H_{12}^{\rm id}$; in reality this becomes a measurement of the actual observable 
$H_{12}$, which will serve to determine whether the encoded qubit is in the
state $\ket{0_{L}} $. We
have thus described a complete scheme for universal QC with the anisotropic
Heisenberg Hamiltonian. Our conclusions remain valid for encodings
into more than three qubits \cite{Bacon:99aKempe:00,WuLidar:unp}. Finally, we note that a dressing transformation can
also be found for the case of QC in the \emph{anisotropic} XXZ model,
$H_{kl} = J_{kl}(S_{k}^{x} S_{l}^{x} + S_{k}^{y} S_{l}^{y}+\delta S_{k}^{z}
S_{l}^{z} + \Delta)$, $\Delta = S_{k}^{x} S_{l}^{y} -S_{k}^{y}
S_{l}^{x}$, $\delta \neq 0$, in the presence of non-uniform Zeeman splittings
\cite{WuLidar:unp}.

{\it Non-separable dressing transformation.---} So far we have
discussed only separable dressing transformations
$V=\bigotimes_{j=1}^{N}V_{j}$. As illustrated by the following simple
example, a non-separable dressing transformation
$V$ may be used to deal with problems such as a one-qubit operation
accompanied inherently by weak two-qubit coupling. Given $N$ qubits, suppose one can
turn on $S_{k}^{z}$ and $S_{k}^{z}S_{l}^{z}$ perfectly, while turning
on $f^{y}S_{k}^{y}$ induces a small inherent error $f^{y}\delta
(S_{k}^{x}{S_{k+1}^{z}}+S_{k}^{x}S_{k-1}^{z})$, with ${\bf
S}_{N+1}={\bf S}_{1}$. This error can be approximately eliminated by a
non-local dressing transformation $V=\exp(i\delta \sum_{k=1}^{N}
S_{k}^{z} S_{k+1}^{z})$ if $\delta \ll 1 $, since $ S_{k}^{y}\approx
V[ S_{k}^{y}+\delta ( S_{k}^{x} S_{k+1}^{z} + S_{k}^{x}S_{k-1}^{z} )] V^{\dagger }$.

\textit{Conclusion.---} We have introduced a general method,
``dressed qubits'', that eliminates arbitrarily strong inherent errors in QC proposals, without
introducing any encoding overhead. Such errors arise when the actual
Hamiltonian driving the system differs from the desired one. Two important
physical examples we have discussed in detail illustrate the power of the
method: elimination of off resonant transitions in multi-level systems in
which a qubit is embedded, and elimination of the inherent spin-orbit
induced anisotropy accompanying the Heisenberg interaction in spin-based QC
proposals.

\textit{Acknowledgements.---} D.A.L. acknowledges support from D-Wave
Systems, Inc., and under the
DARPA-QuIST program (managed by AFOSR under agreement
No. F49620-01-1-0468). We thank the staff at D-Wave
Systems, Inc. for valuable comments.


\end{document}